\documentclass[aps, superscriptaddress ,twocolumn,prl]{revtex4}
\usepackage{amsmath}
\usepackage{amssymb}
\usepackage{epsfig}
\usepackage{graphics}
\usepackage{bm}

\newcommand{\be}{\begin{equation}}
\newcommand{\ee}{\end{equation}}
\newcommand{\bea}{\begin{eqnarray}}
\newcommand{\eea}{\end{eqnarray}}

\newcommand{\rr}{{\bf r}}

\newcommand{\rme}{{\rm e}}

\begin{document}
\title{Spin diffusion in trapped clouds of cold atoms with resonant interactions}
\author{G.\ M.\ Bruun}
\affiliation{Department of Physics and Astronomy, University of Aarhus, Ny Munkegade, DK-8000 Aarhus C, Denmark}

\author{C.\ J.\ Pethick}
\affiliation{The Niels Bohr International Academy, The Niels Bohr Institute, Blegdamsvej 17, DK-2100 Copenhagen \O, Denmark}
\affiliation{NORDITA, Roslagstullsbacken 23, SE-10691 Stockholm, Sweden}

\begin{abstract}
We show that puzzling recent experimental results on spin diffusion in a strongly interacting  atomic gas may be understood  in terms of the  predicted spin diffusion coefficient for a generic strongly interacting system.  
Three important features play a central role: 
a) Fick's law for diffusion must be modified to allow for the trapping potential, b) the diffusion coefficient is inhomogeneous, due to the density variations in the cloud  and c) the diffusion approximation fails in the outer parts of the cloud, where the mean free path is long. \end{abstract}
\maketitle
Diffusion in the presence of an external potential is an important problem in diverse fields, ranging from astrophysics, to condensed matter physics, to biology.  
New vistas for understanding diffusion of spin have been 
opened up by experiments using resonantly interacting atomic gases~\cite{ZwierleinBalanced,ZwierleinPolarized}. 
These experiments are the analog for spin phenomena of earlier groundbreaking experiments that established that  atomic gases may form a perfect fluid with a shear viscosity having the least possible value consistent with   quantum
mechanics~\cite{Cao,Schafer}. 

In the spin transport experiments,  a cloud of  atoms consisting of two hyperfine states of the same atom, which we refer to as $\uparrow$ and $\downarrow$, was studied.   Atoms in one state were displaced with respect to those in the other state and the subsequent dynamics was investigated~\cite{ZwierleinBalanced,ZwierleinPolarized}. 
When the population of one hyperfine species is much larger than the other, the  diffusive motion  is   well described by collisional relaxation~\cite{ZwierleinPolarized,BruunRecati}.  For equal populations 
of the $\uparrow$ and $\downarrow$ atoms, previous studies  have focussed on the initial bouncing motion of the clouds~\cite{Taylor,Goulko}. Here, we analyze the long time scale dynamics and show that, because of the trap potential $V(\rr)$, Fick's law must be modified. 
We demonstrate that this
effect, combined with the fact that the spin diffusion coefficient is inhomogeneous leads to predictions for the decay rate that are more than 1 order of magnitude larger than the experimentally measured one in the classical regime.  The resolution of this puzzle is shown to be the failure of the diffusion approximation
in the outer regions of the cloud. Our analysis 
  accounts for the experimental results in Ref.~\cite{ZwierleinBalanced}
using  the spin diffusion coefficient predicted for a resonantly interacting system. 
There is a rich variety of regimes for spin relaxation, depending
on the trap anisotropy and the density of atoms.

\noindent\emph{Basic formalism\,\,\,}In a trap, the magnetization density $M(\rr)=n_\uparrow(\rr) -n_\downarrow(\rr)$, where $n_i(\rr)$ is the density of species $i$, 
is not constant in equilibrium.  For instance, $M(\rr)\propto e^{-V(\rr)/T}$ for high temperatures  $T$. (We use units in which $k_B=1$.)  Rather the quantity that is constant is the chemical potential difference  $\mu_\uparrow-\mu_\downarrow\simeq 2M/\chi$, where  $\chi=2\partial M/\partial(\mu_\uparrow-\mu_\downarrow)$ is the spin susceptibility.  Thus diffusion is driven by spatial variations of the chemical potentials, and phenomenologically, the spin current density ${\bf j}_M$ is therefore given by the modified Fick's law
\bea
{\bf j}_M=-D\chi \nabla\left(\frac M \chi\right)\label{Fickmod},
\eea 
where $D$ is the spin diffusion coefficient.
Equation (\ref{Fickmod}) reduces to the usual expression ${\bf j}_M=-D\nabla M$ when $V$ is constant.  
We concentrate on the case of temperatures high enough that the gas may be treated using the Maxwell-Boltzmann distribution.  In atomic gases, the dominant 
relaxation process is two-body scattering and, consequently, 
 the diffusion coefficient, which is proportional to the mean free path of a particle, therefore varies inversely with the density $n\propto \rme^{-V/T}$~\cite{ZwierleinBalanced,BruunNJP} and $\chi=n/T$, where $n=n_\uparrow+n_\downarrow$.  

To determine the diffusive modes, we write $M(\rr,t)=e^{-\Gamma t}M(\rr)$.  
Insertion of Eq.\ (\ref{Fickmod}) in the equation of continuity, $\partial_t M+\nabla \cdot {\bf j}_M=0$ gives   
\be
D_0\nabla^2 {P} +\Gamma \rme^{- V/T}{P}=0,\label{Smoluchowski}
\ee
where $D_0$ is the diffusion coefficient at the center of the trap ($V=0$), and $P(\rr) =M(\rr)/n(\rr)$ is the local fractional polarization.
Equation (\ref{Smoluchowski}) describes diffusion in the presence of an external potential and is often referred to 
as the Smoluchowski equation~\cite{Chandrasekhar}.
In regions where $V\gg T$, $P$ satisfies the Laplace equation, and therefore the component of $P$ proportional to the spherical harmonic $Y_{lm}$ must vary as $r^{-l}$ in three dimensions since the solution varying as $r^l$ is forbidden by the condition that, by definition, $|P|\le1$.  Thus both $P$ and $\nabla P$ vanish as $r\rightarrow \infty$. 
Equation (\ref{Smoluchowski}) is therefore analogous to the Schr\"odinger equation for a potential $\propto e^{-V/T}$ and 
determining the eigenvalue $\Gamma$ is  equivalent to finding the strength of the potential that will produce a zero energy bound state. 
Equation (\ref{Smoluchowski}) may  be derived from the variational principle $\delta \Gamma_{\rm var}=0$, where
\be
\Gamma_{\rm var}=\frac{D_0\int d^3r\, (\nabla P)^2 }{\int d^3r\, \rme^{-V/T}P^2} .
\ee
For the lowest mode  with a particular symmetry, $\Gamma_{\rm var}$ provides an upper bound on the lowest eigenvalue. 

\begin{figure}
\begin{center}
\includegraphics[clip=true,width=0.8\columnwidth,height=0.8\columnwidth]{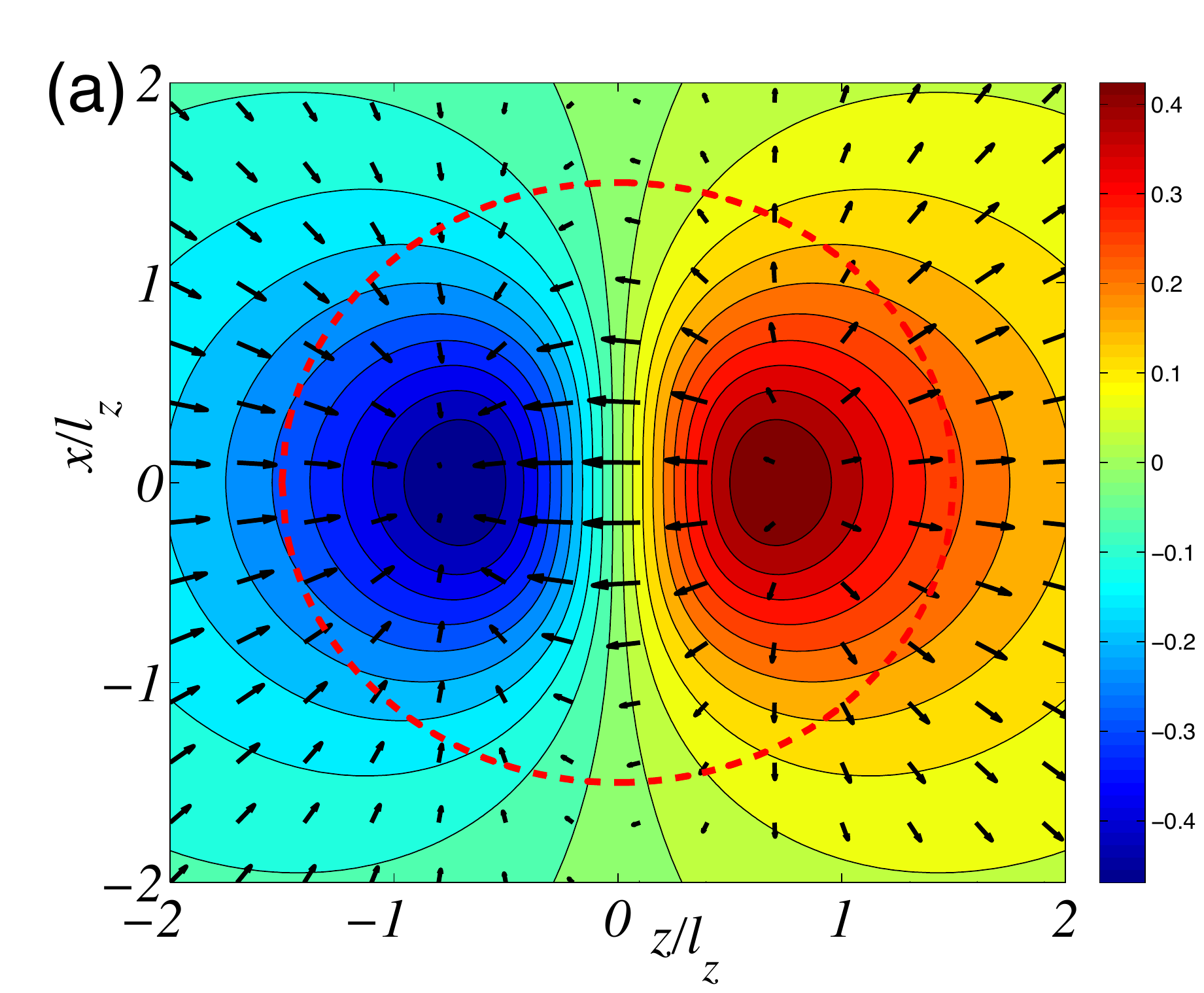}
\includegraphics[clip=true,width=0.8\columnwidth,height=0.8\columnwidth]{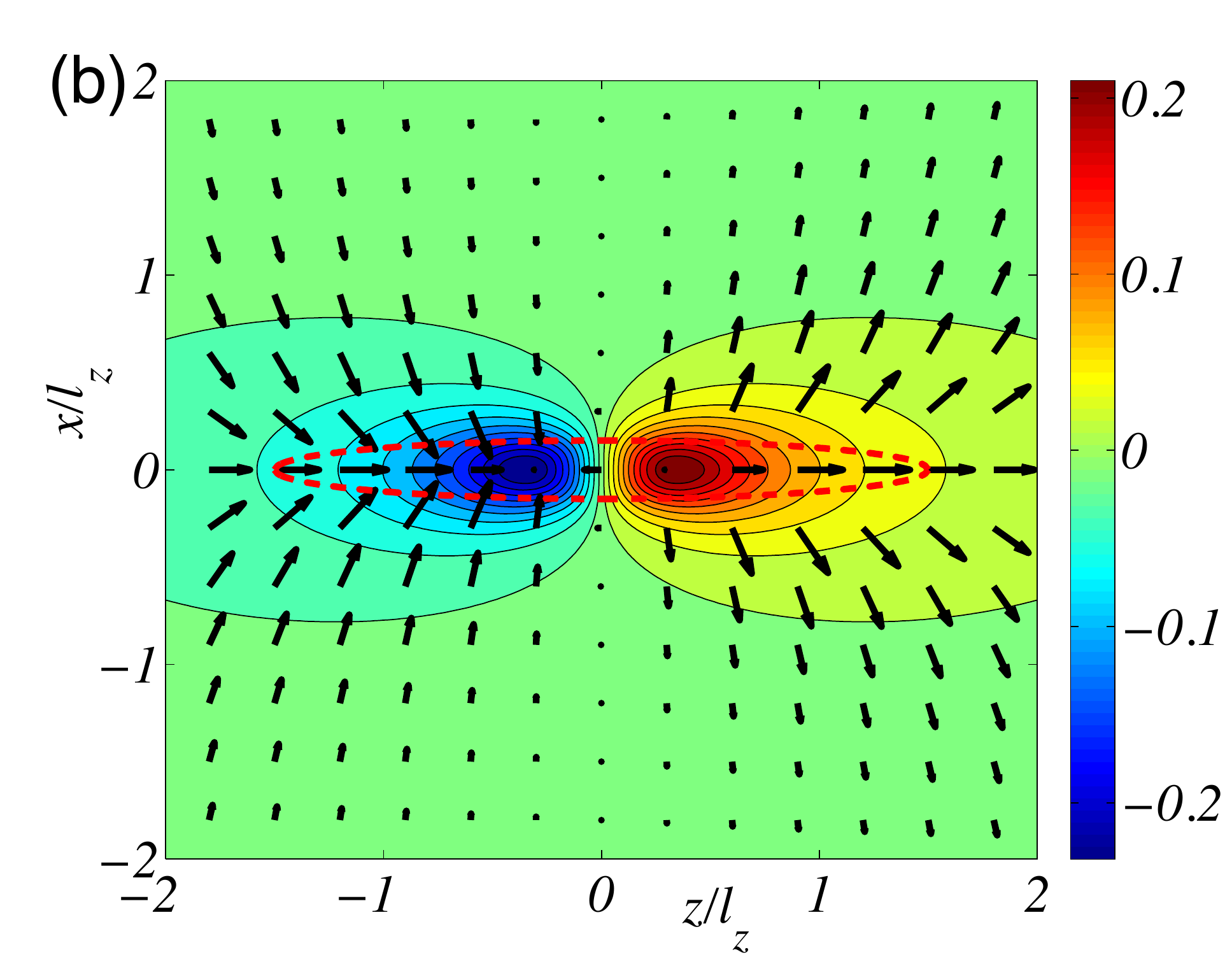}
\includegraphics[clip=true,width=0.8\columnwidth,height=0.8\columnwidth]{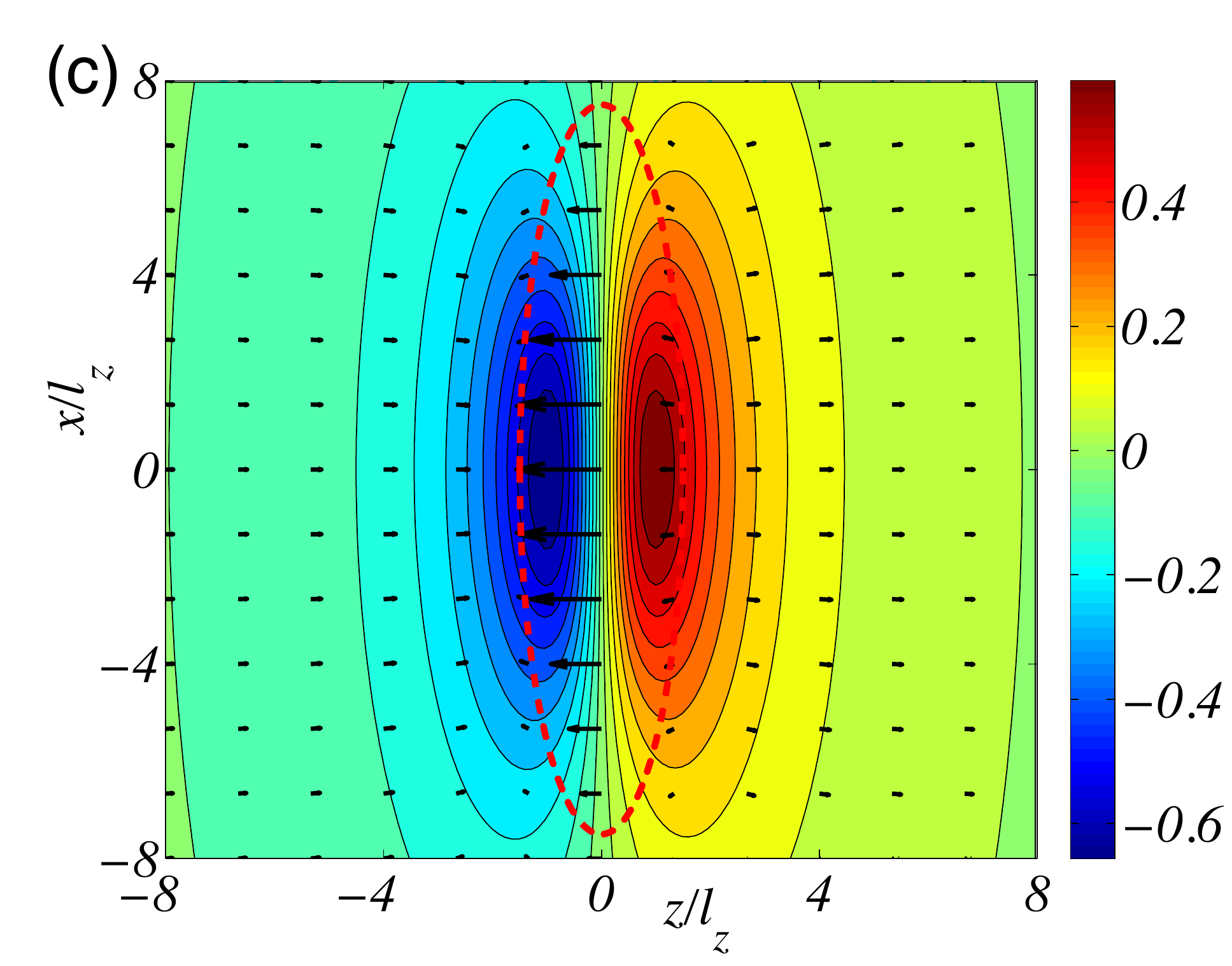}
\caption{(color online) Contour plots of the polarization and  spin current  density (arrows) in the $\rho z$-plane.
 The red dashed contour shows where the density has fallen to 
$0.1$ of the central value. (a) The spherical case. 
(b) The prolate case with $\lambda=\omega_z/\omega_\perp=1/5$.   
(c)  The oblate case with $\lambda=5$.  }
\label{CurrentPlots}
\end{center}
\end{figure}

\noindent\emph{Simple examples\,\,\,}We first solve  (\ref{Smoluchowski}) for a one-dimensional (1D) harmonic potential,
$V=m\omega_z^2 z^2/2$,  where $\omega_z$ is the trap frequency.  
For $|z|\rightarrow \infty$,  $P$ varies as $A+Bz$, where $A$ and $B$ are constants, but because $|P|<1$, $B=0$. 
A numerical solution of  (\ref{Smoluchowski}) for the lowest mode that is odd in $z$ for the boundary condition  $P(z)\rightarrow\text{constant}$
  for $z\rightarrow\infty$  yields for the damping rate the result  
\begin{equation}
\Gamma_{\rm 1D}\approx 2.684\frac{D_0}{l_z^2},
\label{1Dresult}
\end{equation}
 where $l_i^2=2k_BT/m\omega_i^2$. Plots of the polarization and the associated spin current density 
\begin{figure}
\begin{minipage}{.49\columnwidth}
\includegraphics[clip=true,width=0.98\columnwidth,height=1\columnwidth]{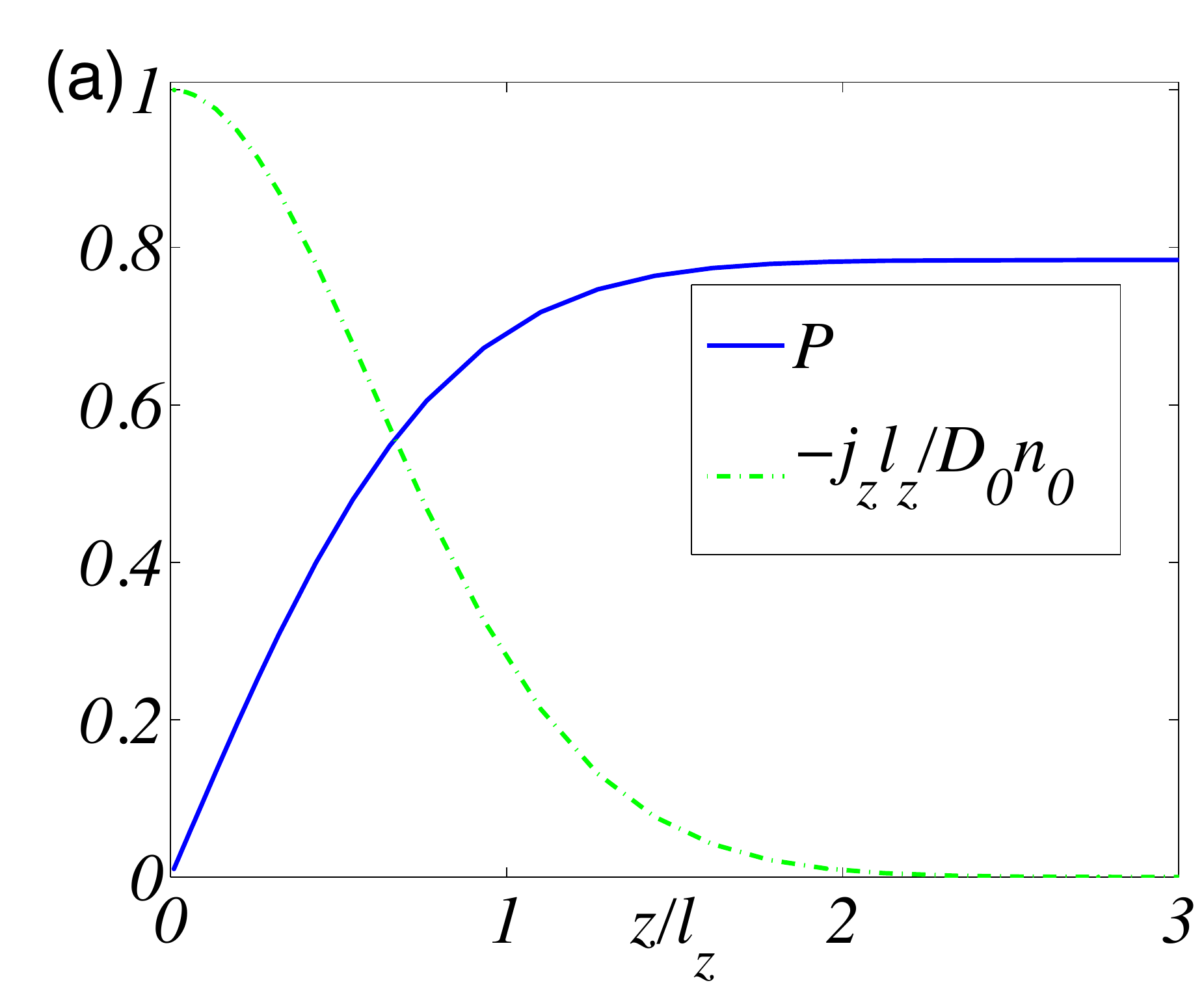}
\end{minipage}
\begin{minipage}{.49\columnwidth}
\includegraphics[clip=true,width=0.98\columnwidth,height=1\columnwidth]{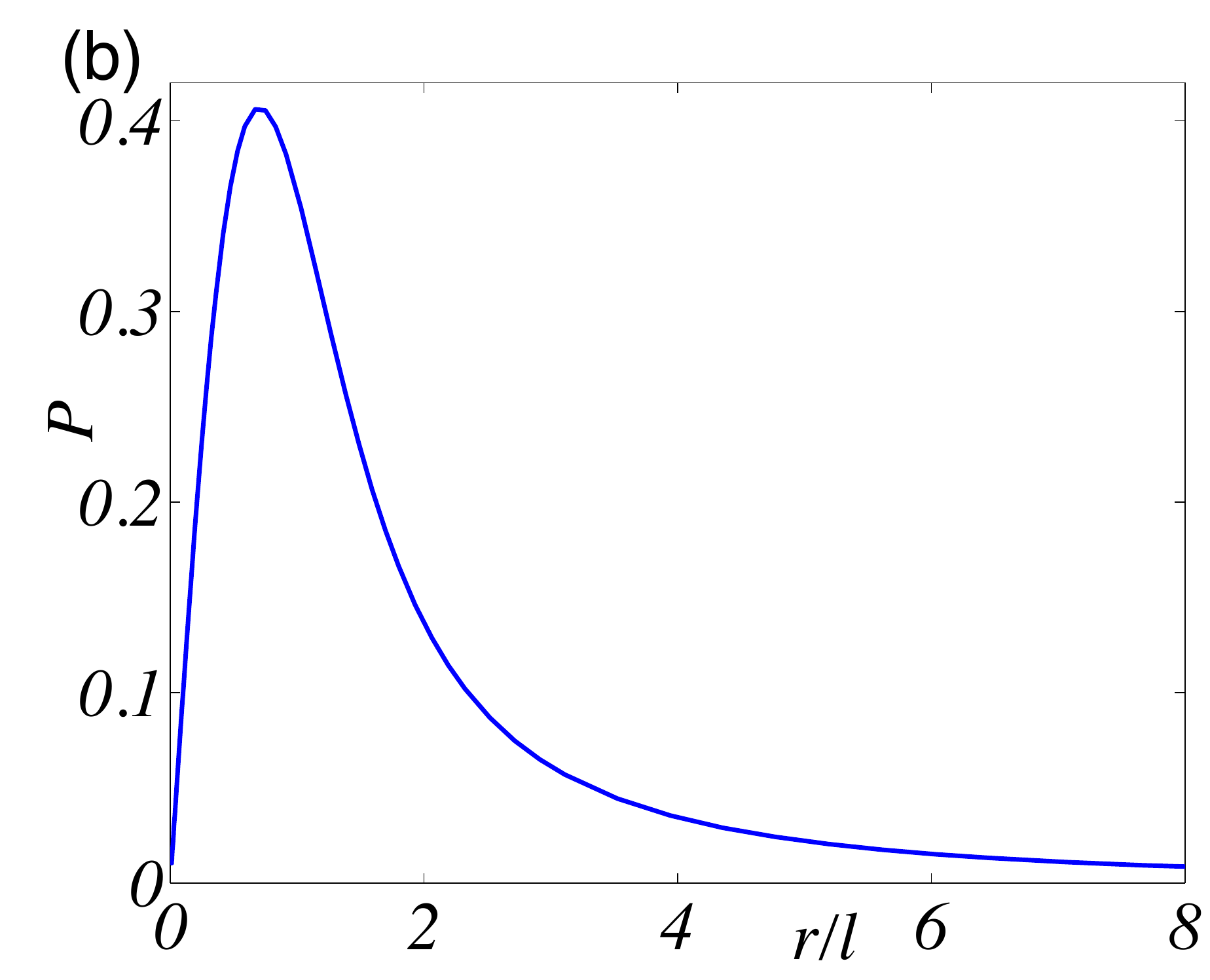}
\end{minipage}
\caption{(color on-line) (a) The polarization (solid line) and the spin current density (dashed line) for the 1D case. 
(b) $P(r)$ for the spherical case with   $P(\rr)=P(r)\cos\theta$.  }
\label{1DandSphericalFig}
\end{figure}
are given in Fig.~\ref{1DandSphericalFig} (a). Since
(\ref{Smoluchowski}) is linear, the normalization
of $P$ and ${\bf j}_M$  in Figs.~\ref{1DandSphericalFig}-\ref{CurrentPlots} is arbitrary.
The variational function $\tanh(z/0.7842 l_z)$ gives  for
$\Gamma_{\rm var}$ the value $2.687 {D_0}/{l_z^2}$, which is within $\sim 0.1\%$  of the exact result.

We now consider the  spherically symmetric case $V=m\omega^2 r^2/2$. 
   The simplest solution rotationally invariant about the $z$-axis and odd in $z$  has the form
$P= Y_{10}(\theta)u(r)/r$ with $Y_{10}(\theta)\propto \cos\theta$.  In Fig.~\ref{1DandSphericalFig} (b), we plot a numerical 
solution to (\ref{Smoluchowski}). Requiring the solution to vanish as $r\rightarrow \infty$ yields the damping rate
 \begin{equation}
\Gamma=12.10\frac{D_0}{l^2}.
\label{ExactSpherical}
\end{equation}
 Figure \mbox{\ref{CurrentPlots} (a)} shows contour plots of $P$ and the spin current density,
  which resembles that for a dipole.  
 The variational function $z/[1+(r/d)^3]$ has the correct asymptotic behavior for both $r\rightarrow 0$ and $r\rightarrow\infty$, and it 
  yields   $\Gamma_{\rm var}=12.12D_0/l^2$ for $d=0.886l$.  

\begin{figure}
\begin{center}
\includegraphics[width=1.0\columnwidth]{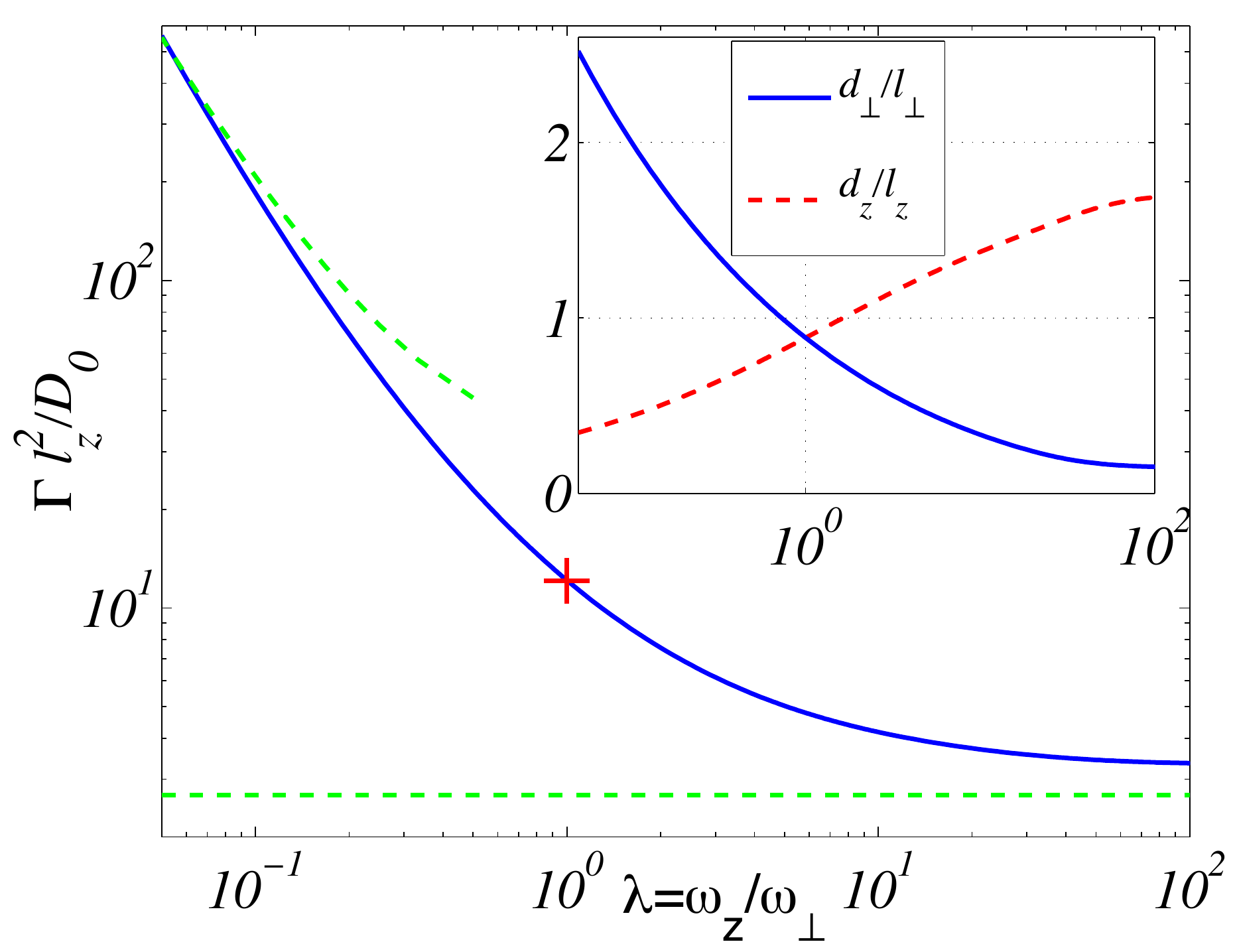}
\caption{(color online) The  damping rate as a function of  the aspect ratio  obtained with the variational function 
(\ref{varNonspherical}) (solid line). The result  (\ref{ExactSpherical}) for a spherical trap   is plotted as a cross, and 
 the limits $\lambda\rightarrow \infty$ (\ref{1Dresult}) and $\lambda\rightarrow 0$  (\ref{DampingCigarEqn}) as dashed lines.  
The variational length scales $d_\perp$ and $d_z$ are shown in the inset. }
\label{Dampvslambda}
\end{center}
\end{figure}

\noindent\emph{Anisotropic traps\,\,\,}The spin diffusion experiments~\cite{ZwierleinBalanced}  are performed in a prolate trap of the form $V(\rr)=m(\omega_\perp^2\rho^2+\omega_z^2z^2)/2=V_\perp+V_z$, 
where $\bm{\rho}=(x,y)$, with $\omega_\perp>\omega_z$. 
We now solve the diffusion equation (\ref{Smoluchowski}) for a general aspect ratio $\lambda=\omega_z/\omega_\perp$ using 
the  variational function 
\begin{equation}
P(\rho,z)= z/(1+\tilde R^3),
\label{varNonspherical}
\end{equation}
with $\tilde R^2= \rho^2/d_\perp^2+ z^2/d_z^2$ which 
obeys the correct boundary conditions for $r\rightarrow0$ and $r\rightarrow\infty$. 
The  variational parameters $d_\perp$  and $d_z$ determine the fall-off of the polarization in the transverse and axial directions in units of $l_\perp$
and $l_z$, respectively. The resulting  damping rate  is 
plotted versus  $\lambda$ in Fig.~\ref{Dampvslambda}.
We see that the  variational function reproduces very accurately the result for the spherical case $\lambda=1$.
 The damping is a decreasing function of $\lambda$, since the transverse confinement imposes a gradient in the polarization as is 
 illustrated in Fig.~\ref{CurrentPlots}.
 The inset demonstrates that for prolate traps, the length scale of the transverse variations $d_\perp$ becomes
  longer than $l_\perp$ while the scale of axial variations $d_z$  becomes shorter than $l_z$; the opposite holds for 
  oblate traps. Figure~\ref{CurrentPlots} (b) illustrates this important point further for the case $\lambda=1/5$:
the polarization distribution is considerably less prolate than the density distribution, and the current density is significant even in regions where the density is low.
Note that, for the prolate and spherical cases, the current has large transverse as well as axial components.
  
  For $\lambda\rightarrow \infty$, we see from Fig.~\ref{Dampvslambda} that the damping rate approaches the 1D result  (\ref{1Dresult}).
 This reflects that the spin motion becomes  1D with the current essentially in the axial direction from the 
 maximum to the minimum of the polarization, as is  clearly seen in Fig.~\ref{CurrentPlots} (c).

 \emph{Born--Oppenheimer approximation\,\,\,} Using the dimensionless variables $\tilde {\bm \rho}=\bm{\rho}/l_\perp$  and $\tilde z=z/l_z$, we see that the diffusion equation (\ref{Smoluchowski}) for an anisotropic, harmonic trapping potential is equivalent to a threshold problem in quantum mechanics with an isotropic 3D Gaussian potential, but where the mass for motion in the transverse directions is a factor 
 $\lambda^2=\omega_z^2/\omega_\perp^2$ smaller than that for axial motion. For the case of a very  prolate trap with $\lambda\ll 1$, we can therefore solve the diffusion equation using the Born-Oppenheimer approximation. Writing $P({\mathbf r})=\psi(z)\phi({\bm \rho},z)$, 
we first find the lowest eigenstate for the ``light" particle by solving 
\be
-\left[\partial_{\tilde x}^2+\partial_{\tilde y}^2-A(z)e^{-\tilde \rho^2}\right]\phi(\rho,z)=E(z)\phi(\rho,z)
\label{Transverseeq}
\ee
with $A(z)=A_0\exp(-\tilde z^2)$ and $A_0=\Gamma l_\perp^2/D_0$;  $\phi(z)$ is determined by solving the equation 
\be
\left[-\frac{\omega^2_z}{\omega_\perp^2}\partial_{\tilde z}^2+E(z)\right]\phi(z)=0.
\label{Axialeq}
\ee
The damping $\Gamma$ is  determined from the value of $A_0$ required for Eq.\  (\ref{Axialeq}) to have a zero-energy bound state that is odd under reflection, $\phi(-z)=-\phi(z)$. To  solve (\ref{Transverseeq}), we observe that for $\omega_\perp\gg\omega_z$  we expect $\Gamma\ll D_0l_\perp^{-2}$ ,  which corresponds to 
$A_0\ll 1$. The transverse problem then reduces to finding the energy of the lowest bound state in a shallow 2D Gaussian potential
$V({\tilde \rho})=-A(z)\exp(-\tilde \rho^2)$. Such a state always exists and for $V_0\ll 1$ its  energy is $E=-\kappa\exp(4\pi/V_0)$
with $V_0=\int d^2\tilde \rho V(\rho)$~\cite{LL,Simon}. The prefactor  is $ \kappa=2\exp(-2\gamma+D_1)$ 
 to lowest order in $A(z)$
 where  $\gamma\simeq0.5772$ is Euler's constant and~\cite{Patil}   
\begin{equation}
 D_1=\int d^2{\tilde {\bm\rho}}_1d^2{\tilde{ \bm\rho}}_2\frac{V({\tilde \rho}_1)V({\tilde \rho}_2)}{V_0^2}\ln\left({\tilde{\bm\rho}}_{12}^2/2\right)
  \end{equation}
with ${\tilde{\bm \rho}}_{12}={\tilde{\bm\rho}}_1-{\tilde{\bm\rho}}_2$. The integrals are straightforward to perform, and we find
\begin{equation}
E(z)=-2e^{-3\gamma}e^{-4/A(z)}.
\label{2Deigenvalue}
\end{equation}
For $A_0\ll 1$, we can expand the eigenvalue as $E(z)\simeq-2\exp[-3\gamma-4(1+\tilde z^2)/A_0]$ and when this is inserted in 
(\ref{Axialeq}), we recover the 1D diffusion problem in a Gaussian trap. Using our 1D result (\ref{1Dresult}), 
 the threshold condition for a bound state odd in $z$ in a very prolate trap  becomes 
\begin{equation}
\Gamma e^{-4D_0/\Gamma l_\perp^2}= 2e^{3\gamma}\Gamma_{1D},
\label{DampingCigarEqn}
\end{equation}
which is an implicit equation for the damping rate. For $\Gamma\ll 1$, the solution  is
\begin{equation}
\Gamma\simeq\frac{4}{\ln\left(l_z^2/ 7.6\,l_\perp^2\right) }\frac{D_0}{l_\perp^2}.
\label{DampingCigar}
\end{equation}
 Thus, the assumption  $A_0\ll1$ for $l_\perp \ll l_z$ 
is consistent. However, the numerical factors indicate that the asymptotic 
expression (\ref{DampingCigar}) is a good approximation only for extremely prolate traps. 
Figure~\ref{Dampvslambda} shows that the variational function (\ref{varNonspherical}) accurately recovers the prolate limit of the damping rate 
given by (\ref{DampingCigarEqn}).

 \emph{Failure of the diffusion approximation\,\,\,} For $\lambda \approx 0.1$ the calculations above predict a damping rate of approximately
$200 D_0/l_z^2$, while experimentally the results of Ref.\ \cite{ZwierleinBalanced} with the expression for $D_0$ from Ref. \cite{BruunNJP} give $\Gamma\approx 10D_0/l_z^2$.  We now demonstrate that the discrepancy is due to the failure of the diffusion approximation in the outer parts of the cloud, where 
 the density is low and conditions are collisionless. 
An approximate  expression for the distance $r_0$ from the $z$ axis at which the diffusion approximation fails may be obtained by arguing that this occurs when a particle has a probability of $1/\rme^\alpha$ of not suffering a collision when it comes in
from infinity.  Assuming$r_{0}\gg l_\perp$, this gives~\cite{Kavoulakis}
\begin{equation}
r_{0}^2\simeq l_\perp^2\ln\left(\frac{n_0\bar\sigma}{2\alpha}\sqrt{\frac{k_BT}{2m\omega_\perp^2}}\right).
\label{r0}
\end{equation}
Equation (\ref{r0}) is correct  to logarithmic accuracy. The exact value of the parameter $\alpha\sim{\mathcal{O}}(1)$   
 may be determined by solving  the kinetic equation in the vicinity of the boundary. However, provided 
$r_{0}\gg l_\perp$ the uncertainty in $\alpha$ has little effect on the value of $r_0$.

At  $\rho=r_0$ it is necessary to impose a boundary condition.  The flux of atoms in the $\rho$-direction for $\rho > r_0$ is small since, in the absence of collisions, atoms moving to larger values of $\rho$ will be reflected by the trapping potential thereby strongly reducing the net flux.  Consequently, we impose the condition that the current in the $\rho$-direction vanish at $\rho=r_0$, or $\partial P/ \partial \rho|_{\rho=r_0}=0$.
When the distance to the boundary $r_0$
 is not much larger than the typical length scale $l$ for the spin diffusion modes, it 
 influences the damping rate. 
In a prolate trap for which $r_0 \gtrsim l_\perp$, the polarization reaches its asymptotic for $z\rightarrow \infty$ for $|z|\lesssim l_z$ and consequently the failure of the diffusion approximation for the motion in the axial direction, which will occur only for distances large compared with $l_z$, does not affect the results. With the boundary condition $\partial P/ \partial \rho|_{\rho=r_0}=0$, 
the variational 
principle derived earlier applies except that the region of integration is limited to   $\rho\leq r_0$.  
For $r_0$ not too much larger than $l_\perp$, we argue that a good approximation for a trial function is simply a function of $z$, which gives a   damping rate  \begin{equation}
\Gamma\simeq \frac{r_{0}^2}{l_\perp^2}\Gamma_{\rm 1D}.
\label{Damping1Dlimit}
\end{equation}

With increasing $r_0$, there will be a  cross-over between the 1D spin currents with a damping given by  (\ref{Damping1Dlimit}) 
 and the fully 3D hydrodynamic spin currents which have both transverse and axial components as shown in Fig.~\ref{1DandSphericalFig} (a)-(b) with a damping 
 scaling as $\Gamma\sim D_0/l_\perp^2$.  We expect the cross-over between 
the two hydrodynamic solutions to occur when $r_{0}^2l_\perp^{-2}\Gamma_{\rm 1D}\sim D_0/l_\perp^2$ which gives  $r_0\sim l_z$.

\emph{Comparison with experiment\,\,\,}We finally compare our results with the experiments in Ref.~\cite{ZwierleinBalanced}. For long times, the spin dynamics is determined by the lowest diffusive mode and the observed decay time is related to the damping rate by the relation $\tau=1/\Gamma$.
Using typical experimental numbers reported  in Ref.~\cite{ZwierleinBalanced}, we obtain  $3\lesssim r_{0}^2/l_\perp^2\lesssim 6$.  
This means that the spin dynamics is diffusive in a major part of the cloud, and since $l_z/l_\perp\simeq10>r_{0}/l_\perp$ we expect the 
motion to be mainly along the $z$-direction with a damping rate given by (\ref{Damping1Dlimit}).
 In Ref.~\cite{ZwierleinBalanced}, the decay time is written in terms of a spin drag coefficient as
  $\tau=\Gamma_{\rm sd}/\omega_z^2$ and we find
\begin{equation}
\frac{\hbar\Gamma_{\rm sd}}{E_F}=2\frac{l_\perp^2}{r_{0}^2}\frac{D_0}{\Gamma_{\rm 1D}l_z^2}\frac{\hbar}{mD_0}\frac T{T_F}
\simeq\frac{l_\perp^2}{r_{0}^2}0.7 \sqrt{\frac{T_F}{T}}
\end{equation}
where we have used (\ref{Damping1Dlimit}) and $D_0\simeq1.1(T/T_F)^{3/2}\hbar/m$ for a strongly interacting  Fermi gas in the classical regime~\cite{ZwierleinBalanced,BruunNJP}. 
This agrees with the measured high temperature experimental result, $\Gamma_{sd}=0.16E_F\hbar^{-1}\sqrt{T_F/T}$ when $r_{0\perp}^2/l_\perp^2\simeq4.4$
which is consistent with the estimate above.  In addition to reproducing the magnitude and temperature dependence of the damping, our 
 result also explains the observation that $\Gamma_{sd}$ is independent of the axial trapping 
frequency $\omega_z$.

In summary, we have shown that  a quantitative account of the measured damping rates of diffusive modes can be given if three novel features of spin diffusion in a trap 
are taken into account: the failure of Fick's law, the inhomogeneity of the diffusion coefficient, and the failure of the diffusion approximation in the outermost regions of the cloud.  The work may be extended in a number of 
 directions: to gases with unequal numbers of the two species, and to degenerate gases, including ones with a condensate of paired fermions.  Calculations of damping rates
may also be refined by improving variational functions and by obtaining a more  quantitative understanding of the boundary between diffusive and collisionless behavior  on the basis of the Boltzmann equation.

We are grateful to A.\ Sommer and M.\ Zwierlein for providing us with their experimental data and for very useful discussions,  and 
to H.\ Fogedby, Andrew Jackson, and Benny Lautrup for helpful conversations. This work was initiated when the authors were participating in the  Nordita workshop 
``Quantum solids, liquids, and gases''. In addition, CJP is grateful to APCTP, Pohang, for hospitality while the manuscript was being written.

\end{document}